\documentclass[sigconf]{acmart}

\AtBeginDocument{%
  \providecommand\BibTeX{{%
    \normalfont B\kern-0.5em{\scshape i\kern-0.25em b}\kern-0.8em\TeX}}}

\copyrightyear{2021}
\acmYear{2021}
\setcopyright{acmlicensed}\acmConference[CHI '21]{CHI Conference on Human Factors in Computing Systems}{May 8--13, 2021}{Yokohama, Japan}
\acmBooktitle{CHI Conference on Human Factors in Computing Systems (CHI '21), May 8--13, 2021, Yokohama, Japan}
\acmPrice{15.00}
\acmDOI{10.1145/3411764.3445708}
\acmISBN{978-1-4503-8096-6/21/05}

\usepackage{wrapfig}
\usepackage{booktabs} % For formal tables
\usepackage{multirow}
\usepackage{pdfcomment}

\begin{document}

\title[``Warm Bodies'': Animating Dynamic Blood Flow on Photos and Avatars]{``Warm Bodies'': A Post-Processing Technique for Animating Dynamic Blood Flow on Photos and Avatars}

\author{Daniel McDuff}
\email{damcduff@microsoft.com}
\affiliation{%
  \institution{Microsoft Research}
  %\streetaddress{P.O. Box 1212}
  \city{Redmond}
  \state{WA}
  \country{USA}
}

\author{Ewa M. Nowara}
\email{emn3@rice.edu}
\affiliation{%
  \institution{Rice University}
  %\streetaddress{1 Th{\o}rv{\"a}ld Circle}
  \city{Houston}
  \state{TX}
  \country{USA}}

\begin{abstract}
What breathes life into an embodied agent or avatar? While body motions such as facial expressions, speech and gestures have been well studied, relatively little attention has been applied to subtle changes due to underlying physiology. We argue that subtle pulse signals are important for creating more lifelike and less disconcerting avatars. We propose a method for animating blood flow patterns, based on a data-driven physiological model that can be used to directly augment the appearance of synthetic avatars and photo-realistic faces. While the changes are difficult for participants to ``see'', they significantly more frequently select faces with blood flow as more anthropomorphic and animated than faces without blood flow. Furthermore, by manipulating the frequency of the heart rate in the underlying signal we can change the perceived arousal of the character.
\end{abstract}




\begin{teaserfigure}
  \pdftooltip{\includegraphics[width=\textwidth]{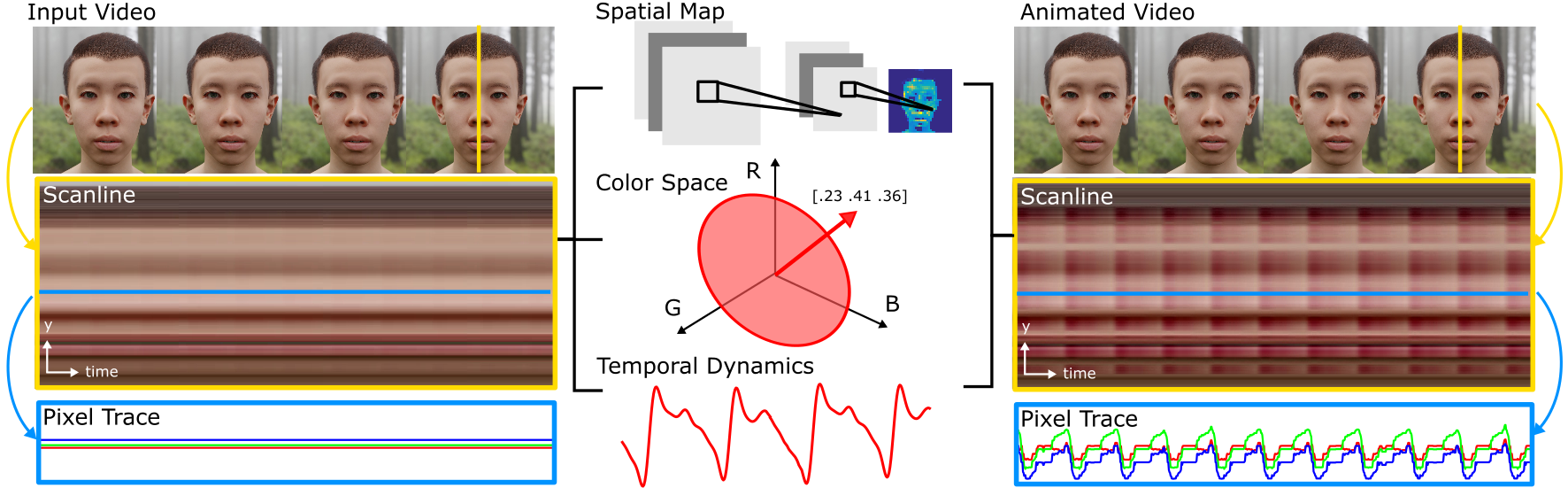}}{A flow diagram showing a visualization of a video before and after augmentation. For both videos (before and after) a sequence of images of a face are shown, below this is a scanline showing spatial slides of pixels across time. After augmentation the scanline shows clear periodic variations.}
  \caption{We present a novel approach to animate subtle physiology directly on virtual characters and photos. Our proposed method incorporates physiologically-grounded spatial, color and temporal modifications into the appearance of the avatar that simulate blood flow. For the purposes of illustration we have magnified the subtle variations (5X) in the vertical pixel scanline (in yellow), the variations in the videos used in our study were much more subtle. We have zoomed in on one pixel within that scanline (horizontal blue line) to show the intensity variations over time in the red, green and blue channels before and after animation.}
  \label{fig:overview}
\end{teaserfigure}

\maketitle
\section{Introduction}

\begin{figure*}
  \pdftooltip{\includegraphics[width=1\textwidth]{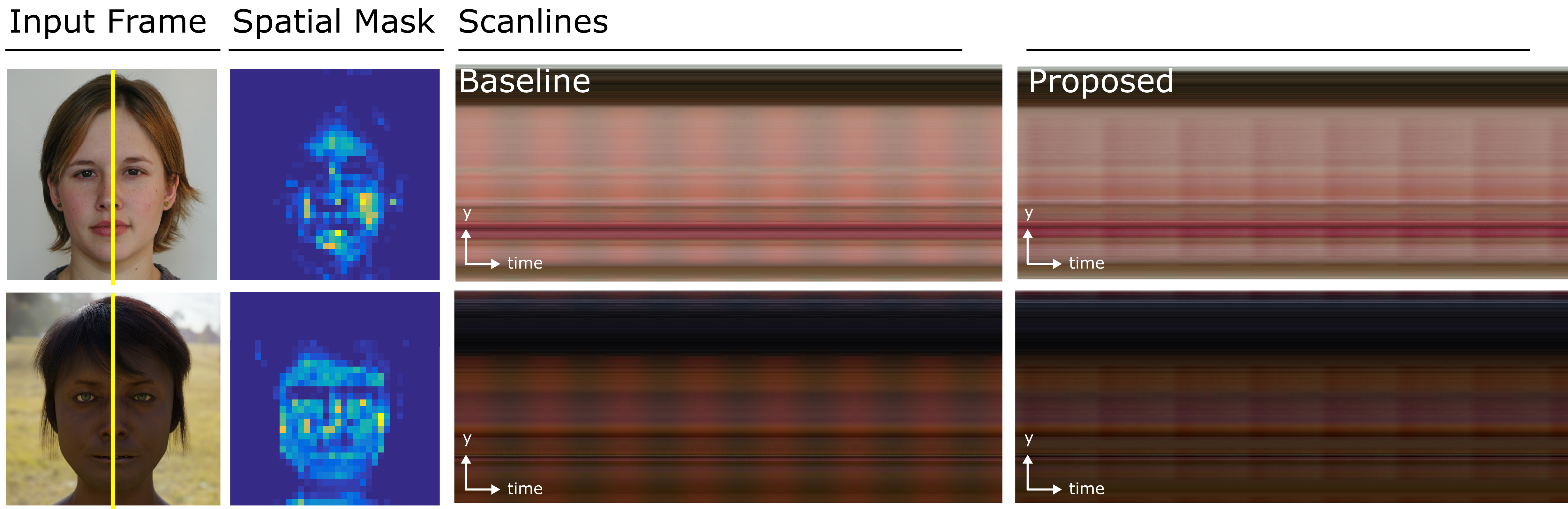}}{Two examples of images of faces, heat maps showing the pulsatile intensity and scanlines. The images shows the difference between augmentation using a baseline and the proposed algorithm.}
  \caption{Scanlines of the animations of two faces: top) real face, bottom) computer generated avatar. Examples of an input image, the corresponding spatial mask from the pre-trained network, scanlines from the baseline animation and proposed physiologically-based animation. For the purposes of these examples we have magnified the changes 5X ($\gamma$=0.75) to make the subtle variations visible, in the user study we used  much more subtle variations ($\gamma$=0.15). The yellow line indicates the vertical pixel slice in each frame used to create the scanline.}
  \label{fig:animationExamples}
\end{figure*}

There are many properties that create believable embodied characters. In Masahiro Mori’s~\cite{mori1970uncanny} graphical representation of the uncanny valley zombies and corpses fall at the very lowest point, this is sometimes known as the ``death mask effect''. What distinguishes an embodied agent as a healthy person versus a zombie, both of which can move, is the appearance that it is fully alive and conscious. Similarly, what distinguishes a video of a person sitting still versus a photograph is subtle motions of the body, such as blinking, breathing or blushing. 
As avatars become more realistic people find them increasingly trustworthy until they are so realistic that small discrepancies become apparent, at this point people can start to find them repulsive~\cite{seymour2019mapping}.

We argue that small differences in the appearance of an agent help create a character that is more natural and lifelike, and that it is important to consider signals as subtle as variations in blood flow. Observers may not be able to ``see'' or identify these changes very easily but we hypothesize that they can still influence their perception and acceptance of an agent. Furthermore, we postulate that even though these changes are subtle, modeling them on real human data creates a more believable effect than using simple heuristics. Even though humans may not be aware of it, we are highly attuned to the nature of even very subtle signals, and can identify when they have unnatural properties.

Using a data-driven approach we animate blood flow on a set of real and synthetic faces (see Fig.~\ref{fig:animationExamples}). We perform a user study to assess how these animations influence perception of the character. Specifically, we focus on how anthropomorphic and animated the faces appear to investigate whether animating blood flow helps combat the ``death mask effect''. There are three main components to our framework (see Fig.~\ref{fig:overview}): 1) spatial mapping of the blood perfusion intensity directly from images or video frames, 2) color space variation based on the absorption profile of hemoglobin and 3) replication and manipulation of temporal dynamics of the pulse (including the systolic and diastolic waveform characteristics). We show that simulating physiological changes on an avatar impacts how people perceive its level of anthropomorphism and animation. Furthermore, simulations that do not follow a physiological model may be rated less well than the original avatar. When avatars are not fully realistic, people can find them repulsive and be reluctant to interact with them. Therefore, creating realistic avatars is important for many human-computer interaction (HCI) systems.

The core contributions of this paper are to: 1) present a novel method of augmenting images with reconstructed dynamic blood flow obtained from a camera and physiologically-based data,
2) show that this can be used to animate faces with physiological changes in a way that makes the person appear more anthropomorphic and animated, 3) investigate which aspects of the augmentation contribute most to the differences in reported judgements about the avatar, and 4) show that manipulating the heart rate can be used to influence perception of the physiological arousal of an avatar. We perform ablation tests to assess which components of the appearance (spatial, color, and temporal) have the greatest impact on the perception of the face. We use open-ended questioning to understand whether viewers could determine the specific changes in appearance of the agent that contributed to their judgements. We argue that animating even subtle changes in physiology is important for creating believable embodied agents and will help these agents from falling into the uncanny valley.

\section{Related Work}
\subsection{Modeling Physiological Changes}

Realistic animation of physiological changes is important for creating characters that look ``alive''~\cite{tsoli2014breathing}. Early research on embodied agents focused on modeling the most obvious behaviors, such as: lip syncing~\cite{taylor2017deep}, head gestures~\cite{jin2019deep}, facial expressions~\cite{kholgade2011content}, gait~\cite{lv2016data,lee2019scalable}, and breathing motions~\cite{promayon1997physically,tsoli2014breathing}. People can easily spot when these behaviors have unnatural intensities or dynamics~\cite{makarainen2014exaggerating}, or when there is a mismatch between artificial and human features~\cite{mitchell2011mismatch,katsyri2015review}. 

Seymour et al.~[\citeyear{seymour2019mapping}] interviewed a group of leading computer graphics experts to gain an understanding of what they viewed as the important challenges involved in creating realistic agents. Blood flow variations in the skin were ``raised as an important issue'' that might make avatars more realistic, even if the changes are difficult for humans to ``see''. Furthermore, Fan et al.~[\citeyear{fan2012real}] found that the appearance of skin was the second most important feature in distinguishing real from computer generated faces. In their study, they were not able to specify what aspects of the skin made it look realistic. We hypothesize that blood flow is one of the properties that is currently missing in synthetic faces.

Advances from the computer graphics community have led to impressive physiologically-based models with multiple translucent layers in the skin~\cite{d2007efficient,jimenez2010real,alkawaz2017oxygenation}. Jimenez et al.~[\citeyear{jimenez2010practical}] presented a dynamic appearance model of skin built from in-vivo measurements of melanin and hemoglobin concentrations. Alkawaz and Basori~[\citeyear{alkawaz2012effect}] presented a system for coloring the face based on the emotional valence of different facial expressions. Other work has investigated skin appearance changes with sweating to create models that capture the effects of physical changes on light reflected from the body~\cite{seo2012rendering}. However, these approaches are not amenable to animating existing photographs or 2-D avatars in a data-driven manner. We present a physiologically-based approach for animating blood flow on faces directly from the video frames without the need for a complex 3D skin model. We confirm the hypothesis of Seymour et al.~[\citeyear{seymour2019mapping}], that blood flow is an important signal and that animating it can help create the appearance of being more alive.

\subsection{Imaging Photoplethysmography}

One reason that little attention has been paid to modeling subtle aspects of human physiology on avatars is the challenging nature of measuring accurate spatial and dynamic patterns of blood perfusion. The peripheral blood flow signal is not spatially uniform, i.e., perfusion is not the same across all regions of skin~\cite{seifalian1994comparison,kamshilin2011photoplethysmographic}. Recent developments in the field of biomedical engineering have led to a suite of techniques termed imaging photoplethysmography (iPPG). Imaging-PPG focuses on methods for recovering and visualizing physiological parameters using everyday cameras and computer vision algorithms~\cite{verkruysse2008remote,poh2010non,mcduff2014improvements,wu2012eulerian,wang2016algorithmic,mcduff2014remote,nishidate2008visualizing,mcduff2018fast,mcduff2017pulse}. 
The development of new computational techniques has led to considerable improvements in the robustness of measurements to large motions, diverse appearances (e.g., skin types), and uncontrolled ambient illumination~\cite{tulyakov2016self,wang2017algorithmic, nowara2018sparsePPG,chen2018deepphys,kumar2015distanceppg}. In early methods~\cite{verkruysse2008remote,poh2010non,wang2016algorithmic}, an aggressive spatial aggregation of pixels was used in order to boost the signal-to-noise ratio before applying a signal processing technique to recover the pulse waveform. As such, they only capture how the intensity of the skin is changing ``globally'' and not local variations. Therefore, they are not well-suited for perfusion estimation on the face.
%However, all of these methods also spatially averaged either an entire facial region or several small regions on the face to reduce noise and to obtain a cleaner one-dimensional pulse signal, and therefore are not well-suited for perfusion estimation on the face.
More recently, spatial maps to model the blood perfusion, which is the rate of change at any tissue region over time, have been proposed. Some of these approaches use a reference signal from a pulse oximeter~\cite{kumar2016pulsecam} or from spatially averaging many pixels over a larger region~\cite{kamshilin2011photoplethysmographic} to map the intensity of the pulse in a given video. The computer graphics community has contributed valuable methods for magnifying the subtle changes of physiology on the body~\cite{wu2012eulerian,wadhwa2013phase,oh2018learning,chen2018deepmag}. However, these methods are only suitable for magnifying signals that are already present in the source material and cannot be easily adapted for animating avatars that are initially static. We use a deep learning approach~\cite{chen2018deepphys} to learn the spatial distribution of blood perfusion color changes on the face. The model generalizes to new faces, and allows us to animate avatars or photos directly from video frames or images which initially have no blood flow signal. 

\begin{figure}
\pdftooltip{\includegraphics[width=0.47\textwidth]{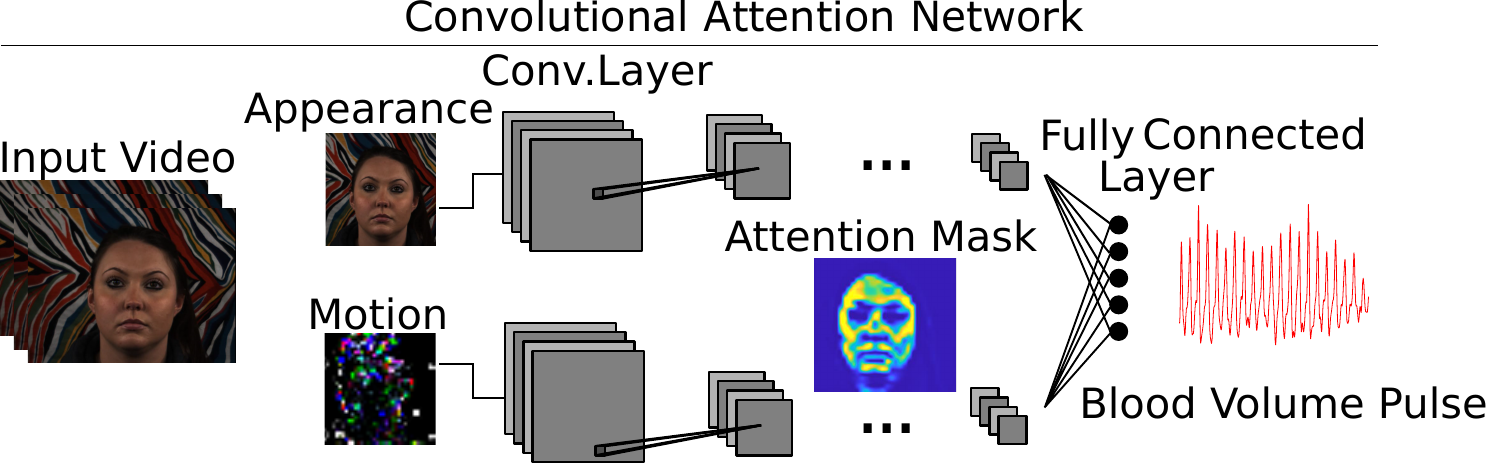}}{A diagram of the architecture of a convolutional neural network architecture with attention mechanism.  The network input is video frames and the output is the predicted pulse signal.}
  \caption{We train a convolutional attention network to recover the blood pulse signal from video. The attention mask shows which regions in the video are likely to contain a strong pulse signal. Once trained we can use the first three layers to recover the perfusion mask for a given frame which captures the pulse signal strength.}
  \label{fig:DeepPhysAdapt}
\end{figure}

\section{Proposed Approach}

We propose an approach for simulating subtle spatio-temporal changes in the appearance of the face of an avatar due to blood flow. Our proposed method captures the spatial distribution, temporal dynamics and color space changes due to blood flow. All three aspects are motivated by physiological data. We animate blood flow on a set of avatars using this approach and compare them to the original avatar and the same avatar animated based on baseline heuristics to simulate blood flow. Our work is the first to use spatial, temporal, and color-based alterations of the appearance of an avatar or photograph to simulate blood flow, therefore there is no obvious other existing baselines. We chose to provide approximations of each component in the formulation of our baseline. For the spatial component we use all skin pixels because only skin regions can contain blood vessels. For the temporal component, we chose to use a periodic sinusoidal signal because the heartbeat is a quasiperiodic signal which can be approximated by a periodic signal on a short time scale. For the color-space component, we chose to use red as the color baseline because the red is the color most commonly associated with blood. The baseline eliminates color changes as a confounding factor, e.g., we test that it is not just that the appearance of the skin is changing at the right frequency but that the changes are modeled in a physiologically accurate manner that matters.

\subsection{Attention Network Architecture}

\label{sec:DeepPhys}
To compute estimates of blood perfusion, we used an existing end-to-end network architecture~\cite{chen2018deepphys}. It uses motion and appearance representations learned jointly through an attention mechanism. The motion branch uses a normalized difference frame and allows the network to distinguish between intensity variations caused by the pulse from those caused by motions. The appearance branch uses the input RGB image to guide the motion representation via the attention mechanism to obtain reliable pulse signals. This joint motion and appearance representation enables the network to be robust to diverse sources of noise including large motions, despite the subtlety of the video-based pulse signal. In addition to a one-dimensional pulse signal estimate we also obtain an attention mask for each video frame. It shows a heat map indicating which regions in the image were used more to compute the pulse, as illustrated in Fig.~\ref{fig:DeepPhysAdapt}. We trained the network on five minute long videos of 25 participants using the contact pulse recordings as labels~\cite{estepp2014recovering}.

\begin{figure}
	\pdftooltip{\includegraphics[width=0.47\textwidth]{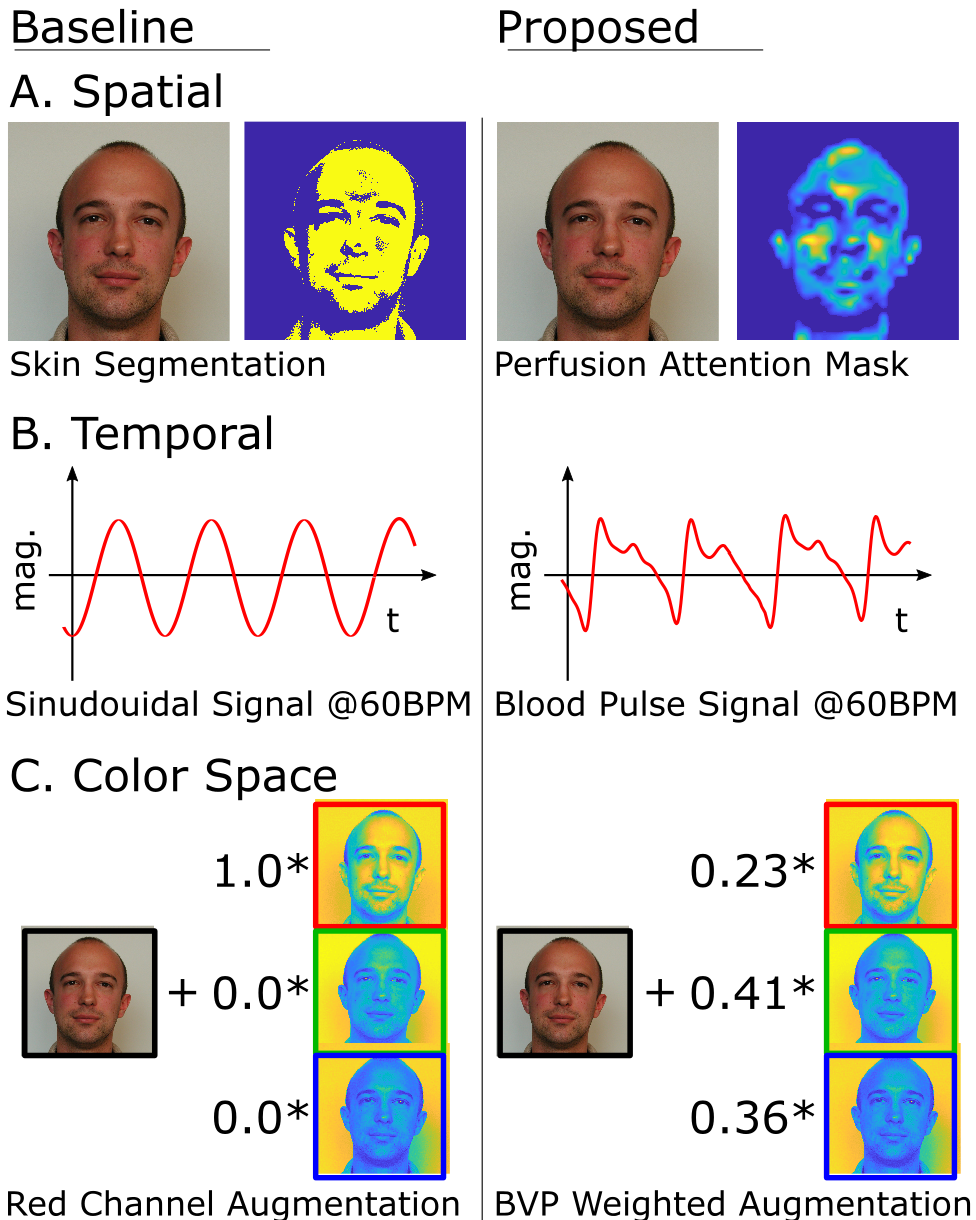}}{A figure illustrating the differences in the spatial, temporal and color space augmentations used in the baseline method and the proposed method for adding physiological variations to an image.}
	\caption{Summary of the face appearance augmentation using our proposed approach based on physiological data and the baseline based on heuristics. To create the augmented videos the final pixel intensities were created by compositing the spatial, temporal and color space weights.}
	\label{fig:baseline_ours}
\end{figure}

\subsection{Spatial Distribution.}
\textbf{Proposed Approach.} 
In order to obtain a spatial map of the blood perfusion intensity we use the attention masks as a prior on the facial blood perfusion. The masks show which regions on the face contain stronger pulse signals. The video augmentation is created by superimposing (adding) attention masks, weighted by a time-varying coefficient to capture the temporal dynamics of the pulse, in an element-wise fashion to the video frames. This is performed for each color channel separately. Fig.~\ref{fig:animationExamples} shows heat maps of attention weights for faces that were not included in the training set to illustrate the learned spatial distribution.

\textbf{Baseline.} As there is no existing baseline for blood perfusion segmentation of this kind, we used skin segmentation as a baseline for the spatial distribution weights. We used an adaptive rule-based skin segmentation method. First, the face was detected in each frame. Next, a histogram of pixel values for each color channel (R, G, B) was calculated for the facial region-of-interest. The modal values for each channel were then used to define thresholds at one half of a standard deviation above and below the mode. 
Pixels that satisfied these thresholds (within half a standard deviation of the modal values) were selected as skin pixels.
Fig.~\ref{fig:baseline_ours}(A) shows a comparison between the mask obtained from our proposed approach and the baseline. As can be seen, the skin segmentation performs well. However, the skin segmentation mask does not contain the nuances of the learned attention mask as the skin segmentation thresholding has no prior information about the spatial distribution of the pulse signal.

\subsection{Temporal Dynamics.}
\textbf{Proposed Approach.} The pulse has a characteristic profile. Each wave typically exhibits a systolic peak and a diastolic peak separated by the dicrotic notch~\cite{brumfield2005digital}. The profile of the pulse varies across the body and therefore we use recordings measured from the face to drive our avatar animations~\cite{estepp2014recovering}.

\textbf{Baseline.} Given the smooth periodic nature of the pulse signal we use a sine wave as a baseline for modeling the temporal dynamics. To create a reasonable baseline the frequency and amplitude of the sine wave were matched to the frequency and amplitude of the pulse used in the proposed method above. Thus, the dominant ``heart rate'' was the same but the waveform dynamics were different in our proposed approach versus the baseline. Fig.~\ref{fig:baseline_ours}(B) shows a comparison between the pulse signal dynamics used for our proposed approach and the baseline.

\subsection{Color Space.} 
\textbf{Proposed Approach.}  
The impact of blood flow on eventual pixel values in a video frame is a function of the illumination and imager used. For our purposes, we simulate just one combination of ambient illumination and imager. We use the findings from a systematic study of the blood pulse signal in image data that revealed the pulse signal strength in the red, green and blue color channel signals to be weighted 0.39, 0.70 and 0.60, respectively~\cite{lin2014using}. These weights were robust regardless of the demixing algorithm applied and support prior work that has found that the green channel has the strongest signal~\cite{mcduff2014improvements,blackford2018remote}. We normalize these weights so that they sum to one. Thus, for our proposed method the final pixel intensity for a frame at time, \textit{t}, is given by:

\begin{align}
    \begin{split}\label{eq:1}
        R'_{xyt} = R_{xyt} + \gamma*0.23*M_{xy}*\texttt{Pulse}_{t}
    \end{split}\\
    \begin{split}\label{eq:2}
        G'_{xyt} = G_{xyt} + \gamma*0.41*M_{xy}*\texttt{Pulse}{t}
    \end{split}\\
    \begin{split}\label{eq:3}
        B'_{xyt} = B_{xyt} + \gamma*0.36*M_{xy}*\texttt{Pulse}_{t} 
    \end{split}
\end{align}

Where $\gamma$ is a tuning parameter that can be used to adjust the strength of the pulse signal. In preliminary experiments we tested the impact of $\gamma$ and found that 0.15 created variations in the avatar that were similar to those in uncompressed videos of real people. The color mapping we used was based on the prior work, which is a weighted combination of the red, green, and blue camera channels which reflects the absorption of hemoglobin in different wavelengths. However, a more sophisticated color map could be learned to account for different lighting conditions and different skin tones.

\begin{figure*}
  \pdftooltip{\includegraphics[width=0.85\textwidth]{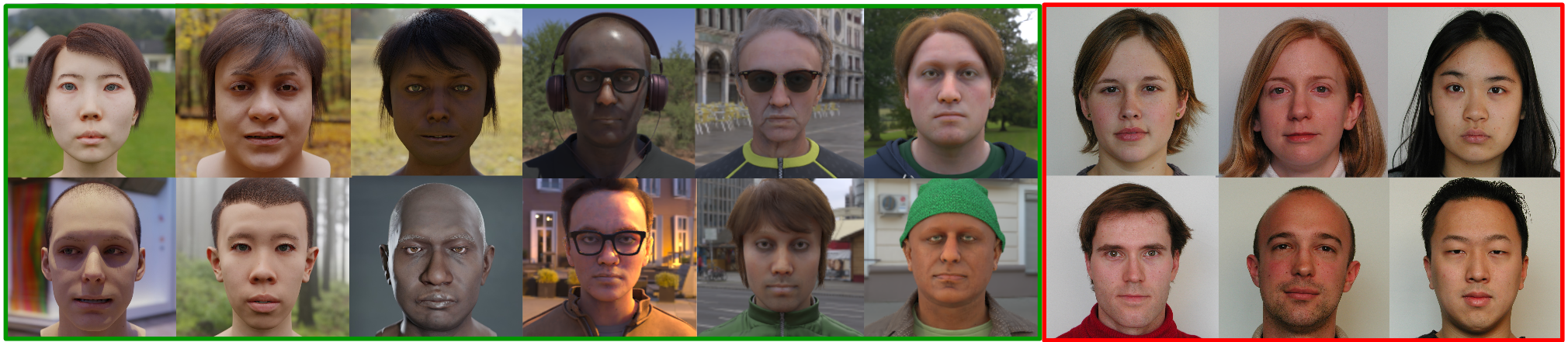}}{Images of eighteen faces, twelve of which are synthetic avatars and six of which are photographs of humans.}
  \caption{The faces we used for animation. Photos of synthetic avatars (green) from an internal dataset and real people (red) from the MIT-CBCL database~\cite{weyrauch2004component}. We intentionally selected a diverse range of skin tones for the faces, with the estimated Fitzpatrick Sun-Reactivity Skin Types of: I-2, II-5, III-4, IV-3, V-2, VI-2.}
  \label{fig:compositeFaces}
\end{figure*}

\begin{figure}
	\pdftooltip{\includegraphics[width=0.47\textwidth]{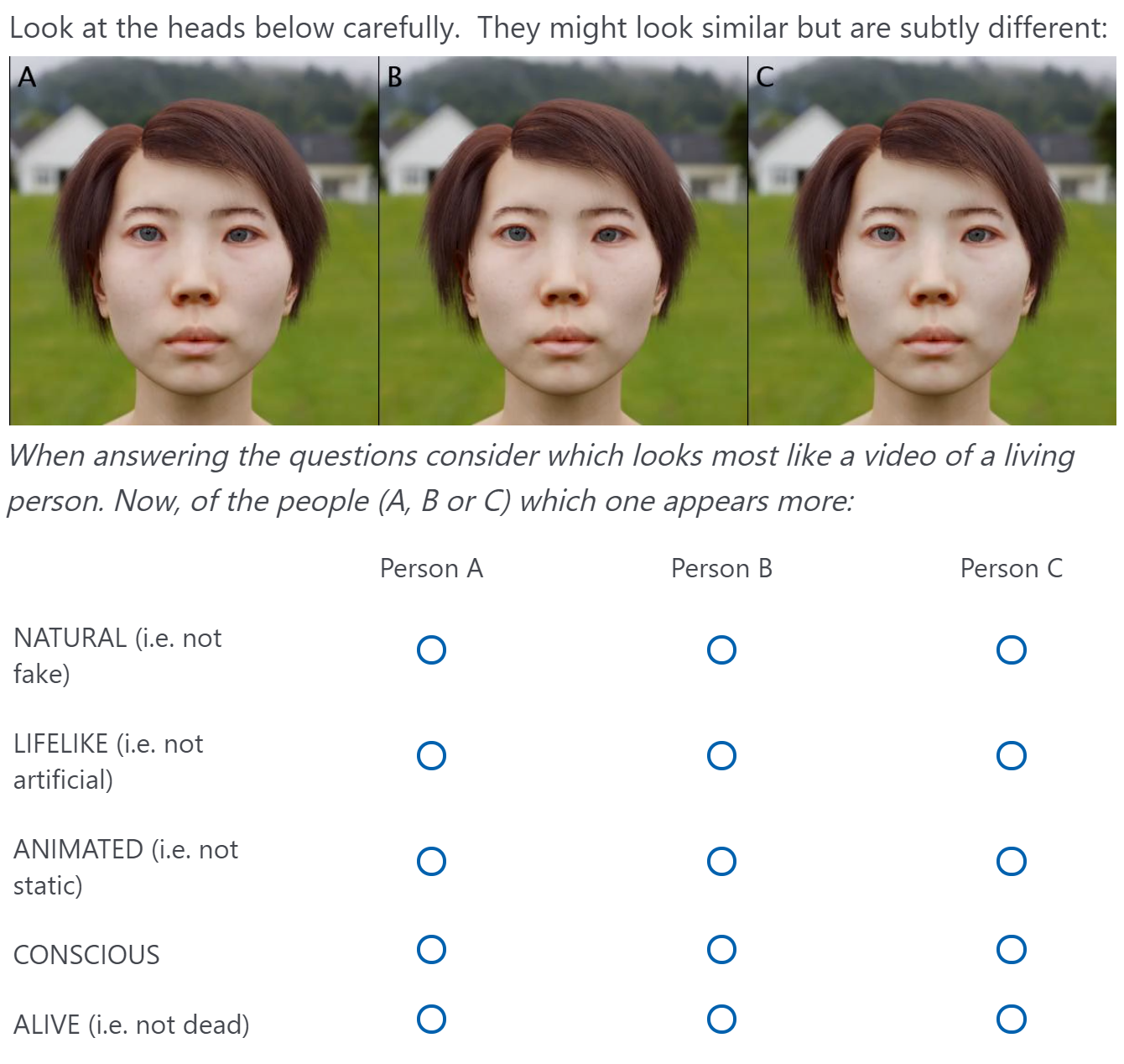}}{A screenshot of an online survey featuring three videos side-by-side and five forced-choice questions.}
	\caption{Screenshot of the human judgement task. Participants watched three videos side-by-side. The order of the videos was randomized. They were asked to make a forced choice about which video featured a person who appeared more natural, lifelike, animated, conscious and alive.}
	\label{fig:questions}
\end{figure}

\textbf{Baseline.} Many people often assume that blood flow influences red coloration of the face the most, because blood is red in appearance. Therefore, our color space baseline is to vary the intensity of the red color channel. Fig.~\ref{fig:baseline_ours}(C) shows a comparison between the spatially averaged RGB values on the face of the augmented avatar obtained from our proposed approach and the baseline.

\section{Experiments}

\textbf{Forced-Choice Perception Judgement.}
We animated two sets of faces (see Fig.~\ref{fig:compositeFaces}) using our approach and the baseline. Twelve faces (3 women, 9 men) were synthetic avatars generated using an in-house set of 3D facial scans. As we expected the effect of the blood flow signal to be small we minimized other motions or behaviors of the avatar to avoid confounding factors or distractions. To simulate adding blood flow to photo-realistic avatars we also animated six (3 women, 3 men) photos taken from the MIT-CBCL face recognition database~\cite{weyrauch2004component}. From these images, we created videos of the faces with simulated blood flow. For simplicity, we will refer to both the synthetic avatars and photos as ``avatars'' going forward. We intentionally selected diverse skin tones for the animated images, with the estimated Fitzpatrick Sun-Reactivity Skin Types~\cite{fitzpatrick1988validity} of: I-2, II-5, III-4, IV-3, V-2, VI-2. Each animation was 10 seconds in length. See the supplementary material for example videos. We recruited 40 subjects 
(27 men, 13 women; mean age: 38.6 years, sd: 9.1 years) on Amazon Mechanical Turk (MT) to watch clips of the avatars in a randomized order. For each avatar participants were shown three animations side-by-side of the same face in the three different conditions. The position (left, middle, right) of the video conditions was randomized, but every question featured the original avatar, the baseline augmentation and our proposed augmentation (see Fig.~\ref{fig:questions}).

\begin{figure*}
	\pdftooltip{\includegraphics[width=1.0\textwidth]{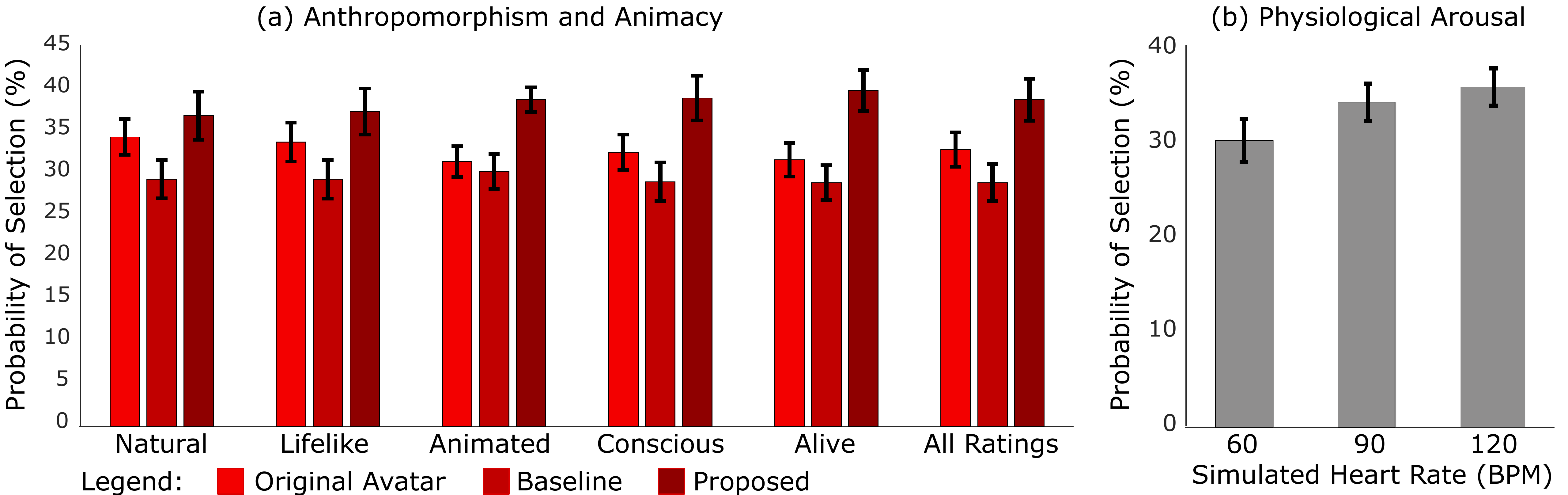}}{Two bar charts. The bar chart on the left is a grouped chart showing three bars in each group and six groups. The y-axis is the probability that a method was selected, for all bars the proposed method was selected more frequently than the other two. The bar chart on the right has three bars.}
	\caption{\textbf{Static Avatar Results (without facial expressions and/or head motions):} Proportion of videos in which each avatar condition was selected as the most: a) natural, lifelike, animated, conscious and alive and b) physiologically aroused. Our proposed approach was selected significantly more frequently overall for the anthropomorphism and animacy metrics. Increasing the simulated heart rate increased the likelihood that the physiological arousal was judged as greater. Error bars reflect the standard error across avatars (N=18).}
	\label{fig:results}
\end{figure*}

To evaluate the clips we adapted questions from the widely used GodSpeed questionnaire~\cite{bartneck2008measuring}. This questionnaire was developed and validated to measure users' perception of robots or avatars. In particular, we evaluated naturalness, lifelikeness, animation, consciousness and aliveness, as these related most to the difference between the avatar being anthropomorphic (natural and lifelike) and animated (animated, conscious and alive). 
Similar to prior work using human subjects to evaluate avatars~\cite{taylor2017deep}, participants were asked to make a forced choice. They were required to chose which person in the videos appeared most: natural, lifelike, animated, conscious and alive. This experimental protocol was approved by the Microsoft Research Institutional Review Board (IRB). Fig.~\ref{fig:results}(a) and Table~\ref{tab:results} show the probability (in percentages) that each method was selected for each question. 

To compare the frequency of selection of the avatars in each condition we used a chi-square test.  In the following analyses we report the Pearson chi-square value ($\chi^2$), number of comparisons (N) and the $p$-value.
Compared to the baseline and the original avatar, our proposed approach was selected most frequently overall 
($\chi^2$ (2, N = 3600) = 52.3, $p$ < 0.01).
Our animation was selected most frequently for naturalness (37\%), lifelikeness (37\%), animation (39\%), consciousness (39\%) and aliveness (40\%).  Individually both naturalness and aliveness were selected significantly more frequently than both the other classes (Naturalness: $\chi^2$ (2, N = 3270) = 26.7, $p$ < 0.01; Aliveness: $\chi^2$ (2, N = 3360) = 26.7, $p$ < 0.01).  For naturalness, lifelikeness, consciousness and aliveness our proposed approach was selected significantly more frequently than the baseline ($p$ < 0.05 for all).

We also performed an evaluation similar to Taylor et al.~[\citeyear{taylor2017deep}], calculating a majority vote for each agent. In this scheme a method ``wins'' the comparison if it receives a majority of the preference judgments from the participants. Ours: 45, Baseline: 19, Original Avatar: 26 ($\chi^2$ (2, N = 90) = 18.9, $p$ < 0.05).
Of the 18 avatars our method won the majority vote for naturalness, animation, aliveness and consciousness. It drew with the baseline for lifelikeness.

At the end of the survey we asked participants what they observed about the avatar that helped them choose between the faces in order to ascertain if they identified the color changes. We did this because the changes were very subtle. Their responses were recorded in an open text field. The participants expressed that they tried to judge the avatars based on how closely they resembled real humans: P4 - \textit{``I answered the questions based on which person in each video appeared most like a real human person and least like a model.''}. However, due to the subtle nature of the animations, many found it difficult to explicitly describe what they based their judgements on.
One participant wrote: P3 - \textit{``They all looked the same to me, so I just relied on my gut instinct to pick the most realistic one.''} and another (P10) wrote \textit{``They all looked exceedingly similar. It was tough to tell a difference.''}, still another (P11) wrote: \textit{``The differences were really subtle and I had a hard time picking out the differences.''} These reflect the most common explanations of ``gut instinct'' or relying on intuition and suggest that in many cases participants did not ``see'' obvious changes in the videos. However, their ratings of the faces were different based on the condition suggesting that the blood flow animations did impact their perception. We might interpret this evidence as meaning that the changes in their perception of the avatars were partially unconscious. In one case a subject noticed the small differences in hue but did not express high confidence: P8 - \textit{``Most of the images looked almost exactly alike, but the slight change in hue was all there was and may have been imagined.''}

\textbf{Real vs Synthetic Faces.} The results were similar for both the synthetic faces and the real faces. In both cases our proposed blood flow augmentation was picked more frequently (37\% of the time across all ratings for synthetic faces and 43\% of the time across all ratings for real faces) than both the original avatar and the baseline augmentation. We interpret this as meaning that blood flow as a subtle signal can be added to both highly (photo-)realistic faces or artificial faces and in both cases help create a more natural appearance.

\begin{figure*}
	\pdftooltip{\includegraphics[width=1.0\textwidth]{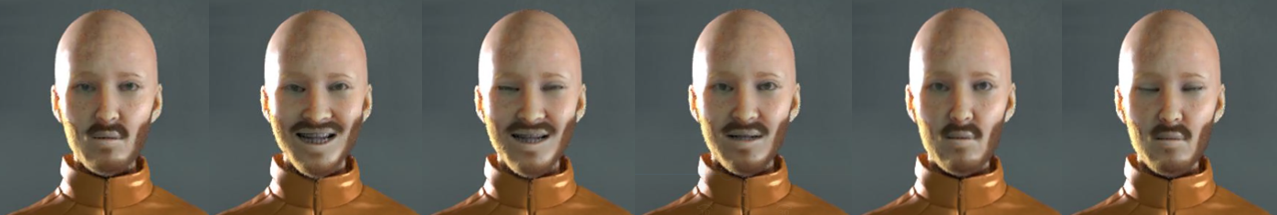}}{Frames from a video of an avatar that is smiling and blinking.}
	\caption{Example frames from a video of an avatar with facial expressions and blinking.}
	\label{fig:videoExample}
\end{figure*}

\textbf{Static vs. Moving Avatars.} In the previous experiments we limited the facial expressions, motions and other appearance changes of the avatars because the physiological manipulation was very subtle. However, it is natural to question whether the effect is still present when there are other sources of variance in the videos. We repeated our experiments by animating blood flow on short (10s) videos of avatars with facial expressions (e.g., smiles), small head motions, and blinking to introduce more natural dynamic behaviors (see Fig.~\ref{fig:videoExample}). Examples of these avatar videos used in our study (and versions with exaggerated/magnified variations - for illustration purposes) can be seen in the supplementary video. We recruited a further 40 subjects (27 men, 13 women; mean age: 37.2 years, sd: 10.9 years) to rate eight of these avatars. 

\begin{figure*}
	\pdftooltip{\includegraphics[width=0.70\textwidth]{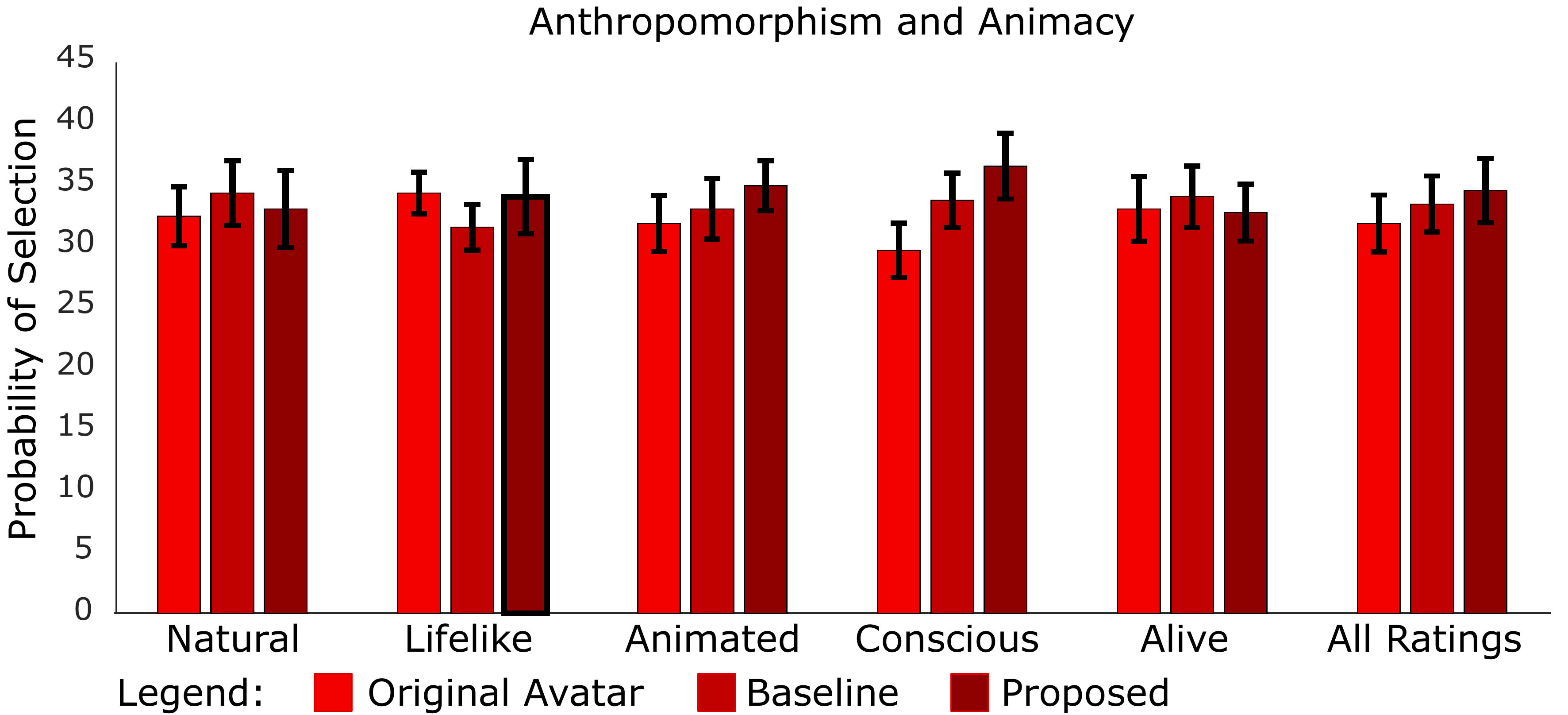}}{A grouped chart showing three bars in each group and six groups. The y-axis is the probability that a method was selected. There is variability in the height of the bars not showing a clear pattern.}
	\caption{\textbf{Moving Avatar Results:} Proportion of videos with facial expressions and head motions in which each avatar condition was selected as the most: a) natural, lifelike, animated, conscious and alive. Error bars reflect the standard error across avatars (N=8).}
	\label{fig:resultsVideo}
\end{figure*}

Overall, we still observe that participants picked the videos with blood flow (proposed) more frequently than videos without blood flow (original avatar) even when there are facial expressions present. However, these differences were not significant based on a chi-square test. The results shown in Fig.~\ref{fig:resultsVideo} and Table~\ref{tab:results} show that there is more variability. This is to be expected and was reflected in the qualitative responses of the participants. %In the presence of larger motions (e.g., facial expressions) the pulse signal 
Many commented on how they judged the videos more on the dynamics of the motions (even though they were identical across conditions). 
In the presence of larger motions and expressions the differences in the skin from frame-to-frame are dominated by the presence of these appearance changes. Therefore, while the participants still often selected the avatars with the physiological animations, there was more variance in the results. Unnatural dynamics in the motion of the avatar or the facial expressions will be apparent. However, we would still argue that as these animations become more realistic and faithful, subtle appearance factors, such as the physiological variations may become increasingly important and noticeable.

\begin{table*}[hbt]
	\caption{Percentage of times that each condition was selected for each of the GodSpeed questions.}
	\label{tab:results}
	\centering
	\small
	\setlength\tabcolsep{3pt} % default value: 6pt
	\begin{tabular}{r|cccccc|cccccc}
	\toprule
	& \multicolumn{6}{c}{\textbf{Static}} & \multicolumn{6}{c}{\textbf{Moving}} \\
        \textbf{Method} & Natural & Lifelike  & Animat. & Conscious & Alive & Overall & Natural & Lifelike  & Animat. & Conscious & Alive & Overall   \\ \hline \hline
        Original Avatar 
        & 34.2 & 33.6 & 31.3 & 32.4 & 31.5 & 32.7
        & 32.5 & \textbf{34.4}  & 31.9 & 29.7 & 33.1 & 31.9 \\
        
        Baseline & 29.2 & 29.2 & 30.1 & 28.9 & 28.8 & 28.8
        & \textbf{34.4} & 31.6  & 33.1 & 33.8 & \textbf{34.1} & 33.5 \\
        
        Proposed & \textbf{36.7} & \textbf{37.2} & \textbf{38.6}  &  \textbf{38.8} & \textbf{39.7} & \textbf{38.6} 
        & 33.1 & 34.1  & \textbf{35.0} & \textbf{36.6} & 32.8 & \textbf{34.6}
\\  \bottomrule
   \end{tabular}
   \\
\end{table*}

\textbf{Ablation Tests.}
To identify which aspect of the avatar's appearance contributed to the differences in the subject's self-reported evaluations we performed an ablation test. We recruited 40 subjects (31 men, 9 women; mean age: 36.8 years, sd: 9.1 years) on Amazon MT. For each avatar we synthesized three videos in the baseline condition but replaced one of either the temporal, spatial or color space augmentations with our proposed approach.  Again, the subjects viewed each avatar side-by-side and were asked to judge which appeared more natural, lifelike, animated, conscious and alive (as in the previous experiment). The color space had the most impact on the judgement of appearance across all categories, with the avatar with our proposed color space weightings being selected as most natural, lifelike, animated, conscious and alive. The temporal and spatial components had similar weights. The avatar with our proposed color space was selected most frequently overall 
($\chi^2$ (2, N = 3600) = 20.9, $p$ < 0.01).

\textbf{Judgement of Affective States.}
The sympathetic nervous system (SNS) regulates a range of visceral functions from the cardiovascular system to the adrenal system~\cite{jansen1995central}. One of the anticipatory responses in humans to an arousing situation is for heart rate to increase. Therefore, we hypothesized that if our simulated blood flow pattern were increased in frequency that people would judge the avatar as being in a state of higher arousal. For each of the 18 static faces we created videos with blood flow (heart rate) frequencies at 60 Beats-Per-Minute (BPM), 90 BPM and 120 BPM. 
Again the position of the video conditions was randomized, but every question featured three versions of the same avatar with a simulated heart rate at 60, 90 and 120 BPM. Participants were asked to chose which avatar appeared most physiologically aroused. We recruited 40 subjects (28 men, 12 women; mean age: 36.1 years, sd: 7.9 years) on Amazon MT who answered this question for all the faces. The probability that a face would be selected increased approximately linearly with the heart rate increase (see Fig.~\ref{fig:results}(b)). This suggests that people interpret the frequency of the blood flow as an indication of physiological state, again reinforcing that it captures appearance change that resembles blood flow reasonably well.
For the purposes of this experiment the magnitude of the blood flow signal was kept constant; however, it is possible that the magnitude of the blood flow signal might also change with arousal level~\cite{nishidate2019rgb}. We will investigate the effects of pulse amplitude variations and vasomotion (the spontaneous low-frequency oscillation in the diameter of blood vessels) in future work.
Again, the participants were asked what they noticed was different about the avatars and how they selected their answers. As in the previous experiments the changes were not obvious and they mostly used their gut instinct or intuition: P10 - \textit{``I just went with my first instinct.''} and P11 - \textit{``I just went with my intuition when making choices.''}. In some cases the subjects thought they saw changes related to particular regions of the face: P9 - \textit{``Eyes were the main thing I used to find the answers .''}, or expressions: P7 - \textit{``I looked at the facial expression and tried to see which one looked more stressed.''}. Again, this suggests that the impact of simulating the blood flow might be somewhat unconscious, but still clearly changed the participant's perception of the physiological state.

\section{Summary}

Physiological processes influence the appearance of the body. We found that subtle augmentation of pixel intensities on an avatar based on physiologically-grounded models of blood flow influences the viewer's perception of the anthropomorphism and animation of an avatar. The avatars (whether photo-realistic or not) were significantly more frequently chosen as anthropomorphic and animated when they had physiological augmentation. 

The qualitative responses of the subjects suggest that most were not aware of the specific manipulations of the avatars (i.e., they could not describe the subtle differences between the conditions) and rather relied on their gut instinct to select their responses. A survey of computer graphics researchers found that some experts thought that blood flow might be an important property for embodied agents, while others questioned whether the changes are even perceptible~\cite{seymour2019mapping}. Our findings suggest that indeed these changes do influence perception of both photo-realistic and computer generated avatars. Given the subtle nature of the differences in appearance, the effects of adding blood flow are modest. This is also presumably because blood flow is only one property of the appearance of an avatar. However, we hope that this work inspires further consideration for adding dynamic blood flow appearance changes to embodied avatars. Our approach allows this to be applied to both synthetic avatars and photo-realistic faces (that might be synthesized using computer vision techniques~\cite{suwajanakorn2017synthesizing}) without the need for a complex 3D graphics model. Our blood flow animation features spatial, temporal and color space components and our experiments reveal that the color space component is the most important for a natural appearance. 

As digital characters become higher fidelity and closer to photo-realistic representations physiological responses will still continue to matter and perhaps become even more important if they help create avatars that traverse the uncanny valley. Adopting animation of blood flow may help avoid the ``death mask effect'' and reduce the potential for people to find faces repulsive.

\section{Limitations and Future Work}
The blood flow signal is very subtle, as we have shown in our work the effects are often unconscious and modestly impact the subject's perception of how anthropomorphic and animated the agent is. We observe the strongest effects in avatars with limited motion but in reality most avatars move (e.g., when speaking or displaying facial expressions). Our preliminary results with animated avatars with smiles, small head movements, and blinking, suggest that the subtle physiological variations may still be important even in the presence of motion. However, further investigation is needed as to how results might be moderated by other activities of the avatars and the effects are likely to still be small. We focused specifically on appearance changes related to blood flow; however, many physiological processes are interrelated. For example, respiratory sinus arrhythmia (RSA) is the phenomenon in which heart rate variability (HRV) is synchronized with respiration. Combining our approach with models for animating other physiological changes (for example breathing, blood oxygen saturation levels and perspiration/sweating) could create even more natural and lifelike representations.

We did not restrict the devices or displays that the participants used in the study. These differences could influence how visible the physiological variations were. However, as our experiment was a within subjects design the three versions of the avatar shown to the user would have been affected similarly. The participants consistently reported that they did not obviously see the intensity variations, regardless of the display they used. Despite not restricting the devices used by the participants in the study, they consistently selected the avatars with physiological variations. Furthermore, it would be very interesting to investigate whether the effects observed in our study are stronger when viewing avatars in virtual reality. Our hypothesis is that they would be.

The focus of our study was to analyze whether animating avatars with physiological variations has an effect on how the avatars are perceived. Future work could consider more open-ended questions and a comprehensive thematic analysis to understand how these augmentations influence the perception of the avatars.

It has been demonstrated that people with darker skin types have weaker iPPG signals because of higher melanin concentration which absorbs more light inside the skin~\cite{nowara2020meta}. While we included faces from a diverse range of skin types in our experiments, we did not perform a systematic analysis of the effect of different races as we did not have a large number of avatars from each skin type category. More work needs to be done to verify that introducing pulse signals improves the appearance of avatars of different ethnicities; and whether the amplitude of these signals should be adjusted to achieve a realistic appearance on avatars with different skin types.

\bibliographystyle{ACM-Reference-Format}
\bibliography{sample-manuscript}

\end{document}